\documentclass[
    ,final            
  ]
  {aipproc}

\layoutstyle{6x9}

\begin{document}

\title{CDF - The Top Quark Forward Backward Asymmetry}


\author{Yen-Chu Chen}{
  address={Institute of Physics, Academia Sinica, Taiwan, ROC}
}

\begin{abstract}
It has been more than 15 years since the discovery of the top quark.
Great strides have been made in the measurement of the top quark mass
and the properties of it. Most results show consistency
with the standard model. However, using 5 fb$^{-1}$ data, recent 
measurements of the asymmetry in the production of top and anti-top 
quark pair have demonstrated surprisingly large values at CDF. Using
4 fb$^{-1}$ data, D0 also has similar effect.
\end{abstract}

\maketitle


\section{Introduction}

The forward-backward asymmetry, ${A_{fb}}$, of the top quark pair production are 
studied by both D0 \cite{D0_Afb_2008} and CDF \cite{CDF_Afb_2008} in 2008. Recently 
the physics of top quark has become very interesting since the finding of the 
forward-backward asymmetry, ${A_{fb}}$, in the top quark production has become 
more significant. 

At leading order of QCD there is no asymmetry in the top quark pair production due 
to the Charge symmetry. The asymmetry can arise only in the next to leading order, 
where interference occurs between the Born amplitude and two-gluon intermediate 
state as well as the gluon bremsstrahlung and gluon-(anti)quark scattering into 
$t \bar{t}$. This asymmetry is predicted to be about $0.078$ \cite{QCD_NLO_1} 
\cite{QCD_NLO_2} \cite{QCD_NLO_3}. Recent calculation based on next next to leading
order shows $0.052 ^{+0.000}_{-0.006}$ \cite{QCD_NNLO}.

At CDF using events having one high transverse momentum lepton and multiple jets, 
the corrected parton level asymmetry is $0.158 \pm 0.074$. Recent analysis using 
the events having two high transverse-momentum leptons and multiple jets shows 
that at the parton level the obtained $A_{fb}$ is $0.42 \pm 0.15 (stat) 
\pm 0.05 (sys)$. Both results show interesting deviation from the theoretical 
predictions. 

The differential cross section of top quark pair is dependent on the production 
angle and the Q value of the production. That means the forward-backward asymmetry 
is dependent on $\Delta y_t$ and $M_{t\bar{t}}$, where $\Delta y_t$ is the rapidity 
difference of the top and anti-top quarks, and $M_{t\bar{t}}$ is the invariant mass 
of the top, anti-top quark pair. Using the lepton plus jets events, for $M_{t\bar{t}} 
\ge $ 450 GeV/$c^2$, the parton level asymmetry is $0.475 \pm 0.114 (stat + sys)$ 
to be compared with the next-to-leading order QCD prediction of $0.088 \pm 0.013$.

\section{The analysis}

The events used in the analysis are collected by the Collider Detector Facility (CDF) 
at the Fermi National Accelerator Laboratory \cite{CDF_Detector_1} \cite{CDF_Detector_2}. 
The relevant components to this analysis are the silicon tracker, the central outer 
tracker, the electromagnetic and hadronic calorimeters, the muon detectors, and the 
luminosity counters.

At the Tevatron the top quark pairs dominantly are produced via quark anti-quark annihilation, 
85\%, and from gluon fusion, 15\%. The top quark decays almost exclusively to a $W$-boson 
and a $b$ quark. Where the $W$-boson can decay leptonically or hadronically. Based on how 
the $W$-boson decays the events are categorized. For the events having only one $W$-boson 
decays leptonically, there are one charge lepton, two light quarks and two $b$ quarks in 
the final state. These are the lepton plus jets events. For the events having both $W$-bosons 
decay leptonically, there are two charge leptons and two $b$ quarks in the final state, 
these are the di-lepton events. In case of both $W$-boson decay hadronically, these are the 
all hadronic events. The all hadronic events are dominated by QCD background and are not 
used in this analysis.

\subsection{Using the lepton plus jets events}

For this analysis the event is triggered by a high transverse momentum electon(muon) 
in the central portion of the detector with $E_{T}(p_T)$ > 20 GeV(GeV/c) and $|\eta|$ < 1.0.
The jets originated from quarks are reconstructed using a cone algorithm with 
$\delta R = \sqrt{\delta \phi^2 + \delta \eta^2}$ < 0.4.
Four or more jets are required with $E_{T}$ > 20 GeV and $|\eta|$ < 2.0. Large missing
transverse energy is required, ${/\!\!\!\!E_t}$, $\ge$ 20 GeV to reduce the QCD background.
To further reduce the background, the SECVTX algorithm is used to find displaced $b$-decay
vertices using the tracks within the jet cones; at least one jet must contain such a vertex. 
From data corresponding to an integrated luminosity of 5.3 fb$^{-1}$ 1260 events pass the
event selection. 

\begin{figure}
  \includegraphics[height=.25\textheight]{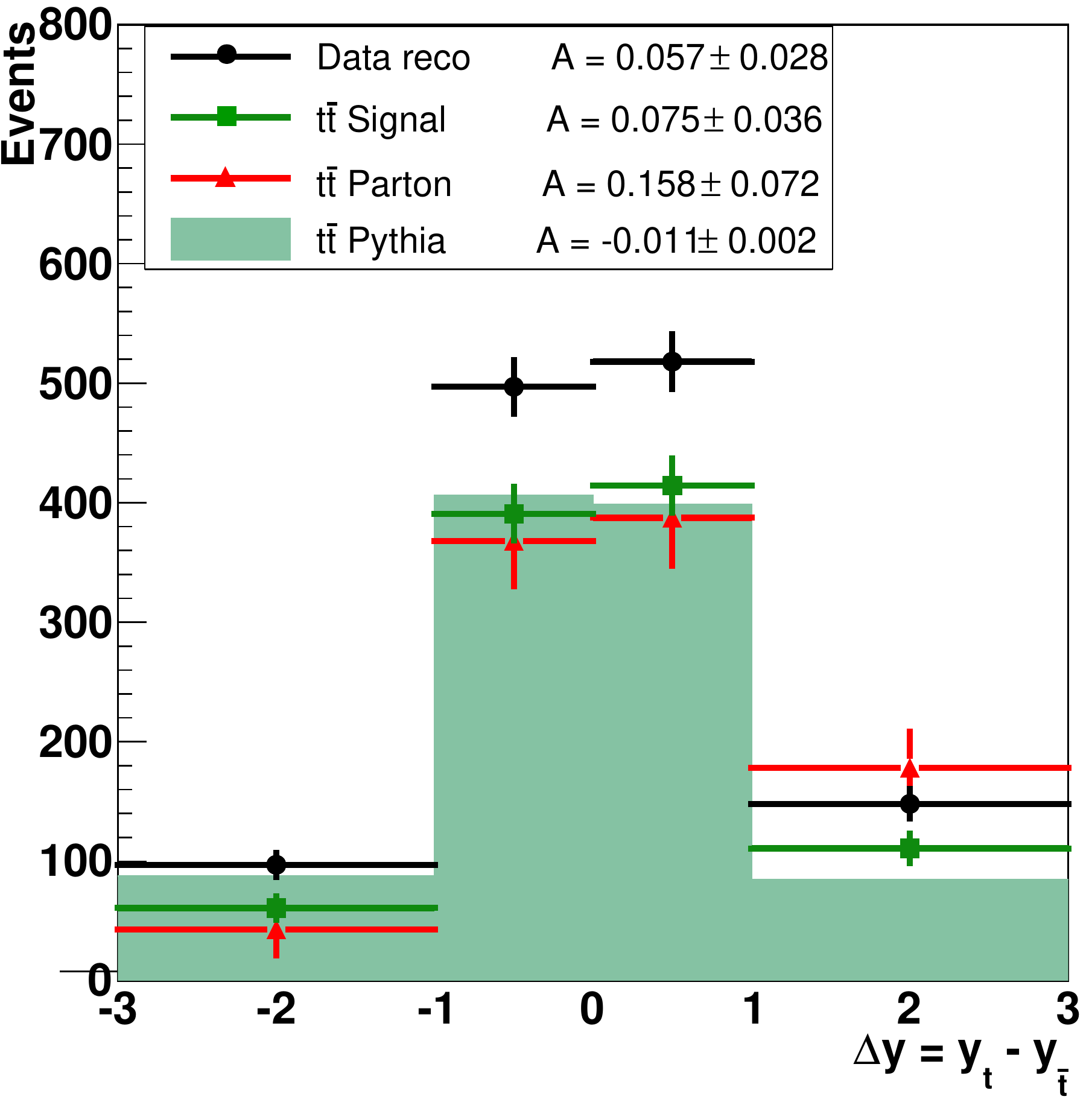}
  \caption{The rapidity difference of top and anti-top quark from candidate events.}
  \label{Afb_LJ}
\end{figure}

Given the lepton, missing $E_{T}$, and four or more jets, possible combinations are tried
to calculate the $\chi^2$, contraint by the top mass and $W$-boson mass. The most probable 
solution is chosen to calculate the rapidity of the top and anti-top quarks, $y_t$ and 
$y_{\bar{t}}$. Distribution of $\Delta y = y_t - y_{\bar{t}}$ is shown in FIGURE
\ref{Afb_LJ}. The observed $A_{fb}$ from CDF is $0.057 \pm 0.028$, where the uncertainty
includes statistical, systematic, and theoretical uncertainties. With the
contribution from the background subtracted this becomes $0.075 \pm 0.037$.
Which is consistent with the measurement from D0, $0.08 \pm 0.04$. 
Taking into account the detector acceptance, reconstruction smearing, the parton level
asymmetry can be obtained as $0.158 \pm 0.072$.

\begin{figure}
  \begin{tabular}{cc}
    \includegraphics[height=.25\textheight]{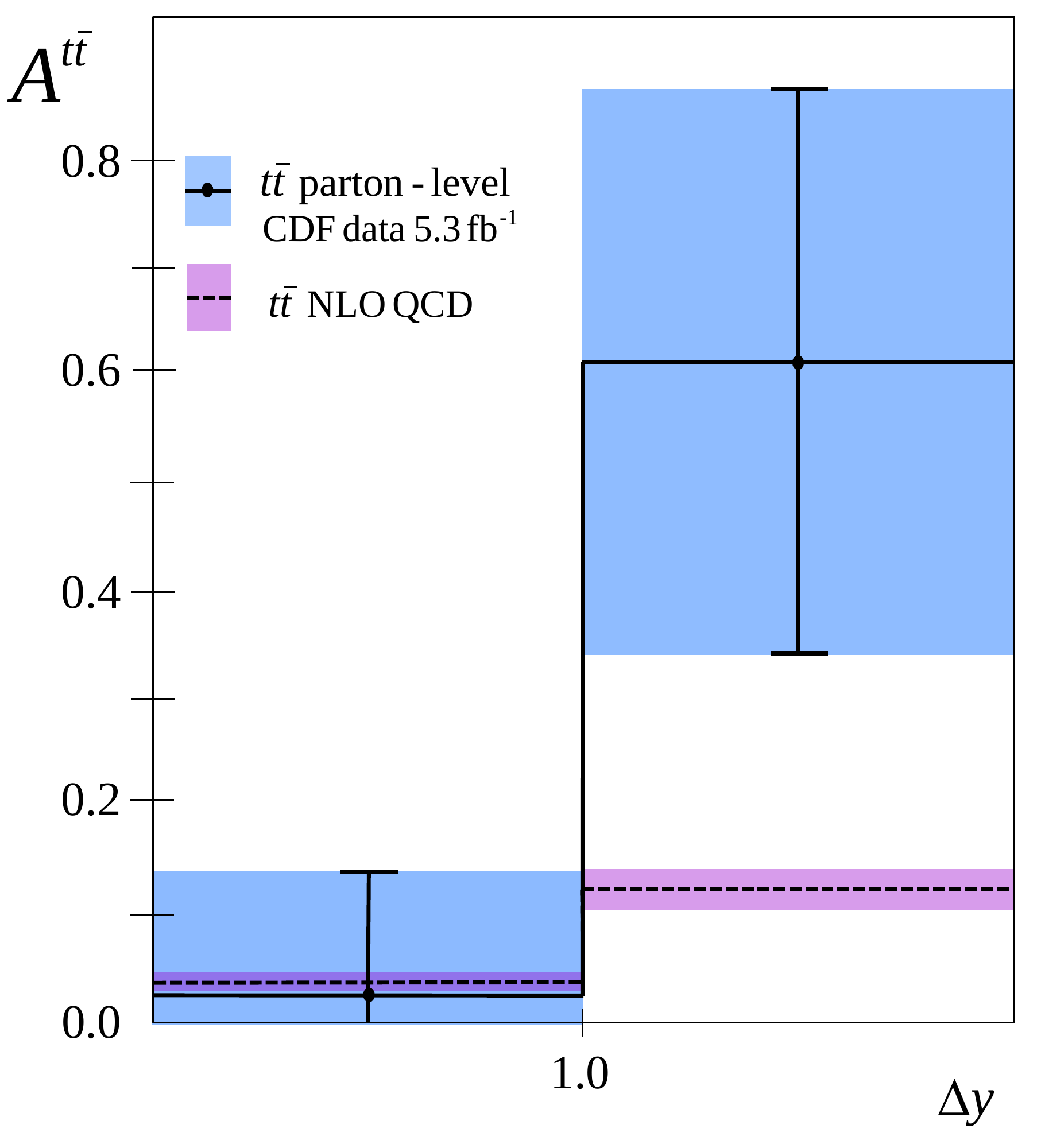} &
    \includegraphics[height=.25\textheight]{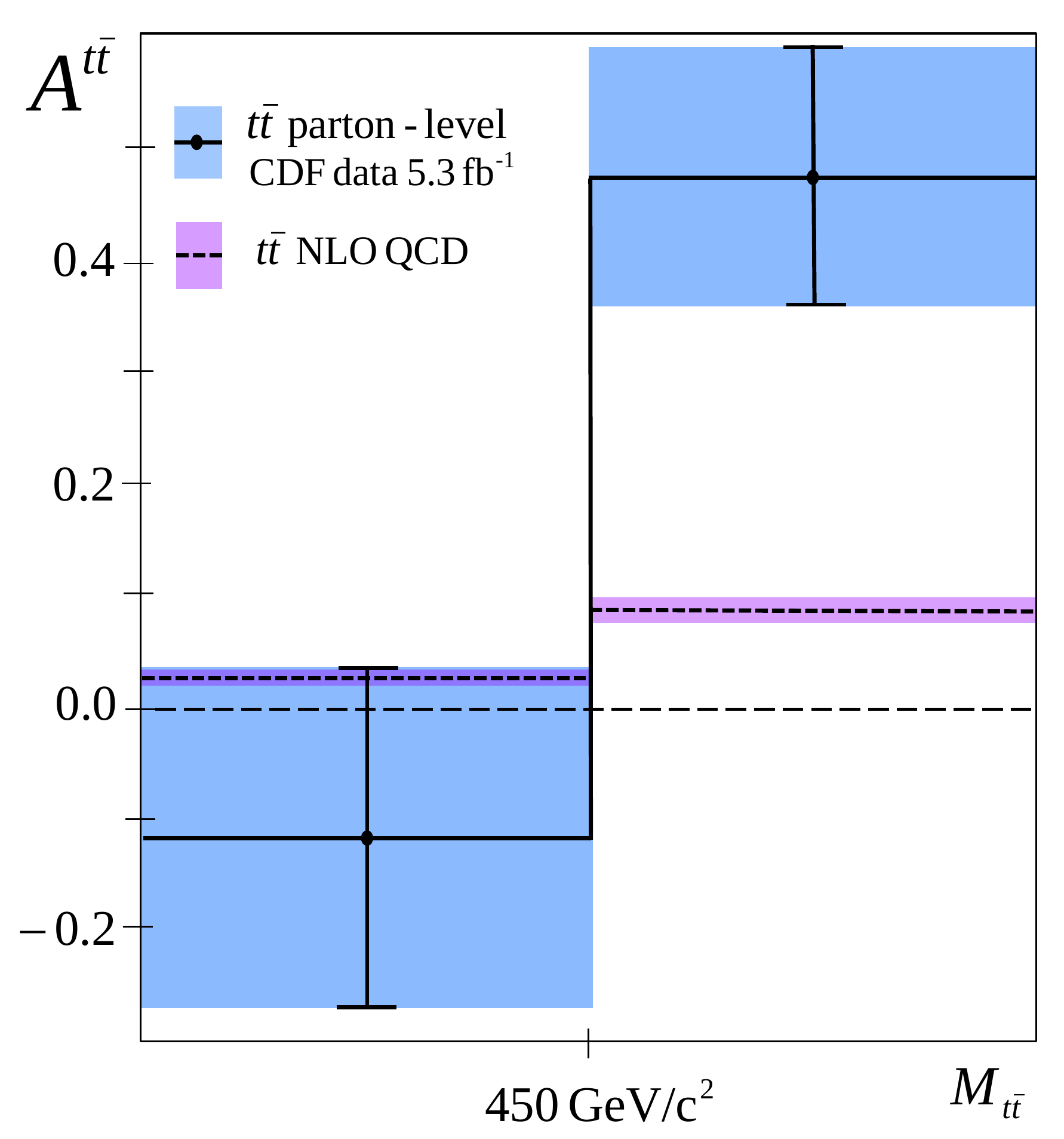} \\
  \end{tabular}
  \caption{Asymmetry dependence on $\Delta y$ and $M_{t\bar{t}}$.}
  \label{Afb_DY_Mtt_LJ}
\end{figure}

The differential production cross section of top quark pair is dependent on the production
angle and the Q value. This means $A_{fb}$ is dependent on $\Delta y$ and $M_{t\bar{t}}$.
FIGURE \ref{Afb_DY_Mtt_LJ} shows the dependence on $\Delta y$, where larger asymmetry is 
observed at larger $\Delta y$. FIGURE \ref{Afb_DY_Mtt_LJ} shows the dependence on the 
$M_{t\bar{t}}$. When $M_{t\bar{t}} > $ 450 GeV/c$^2$ the asymmetry is particularly large, 
$0.475 \pm 0.114 (stat + sys)$; which is not consistent with the QCD NLO prediction, 
$0.088 \pm 0.013$.

\subsection{Using the di-lepton events}

The di-lepton events are selected when there are two leptons with
high transverse momentum, $p_{T}$ > 20 GeV/$c$.
At least one lepton is from the central portion of the detector.
Jets are reconstructed with cone size of 0.4. Two or more jets
with $E_{T}$ > 15 GeV and $|\eta|$ < 2.5, should appear in a given 
event. To reduce the QCD background we require high missing transverse
energy, ${/\!\!\!\!E_t}$ > 25 GeV or ${/\!\!\!\!E_t}$ > 50 GeV if
there is any lepton or jet closer then 20$^0$ with respect to the
${/\!\!\!\!E_t}$. To further reduce the background additional 
requirement is placed on the scalar sum ($H_T$) of the transverse
energy of the leptons, ${/\!\!\!\!E_t}$, and jets, $H_T$ > 200 GeV.
From data of 5.1 fb$^{-1}$ 334 events pass these criteria.

\begin{figure}
     \begin{tabular}{cc}
        \includegraphics[height=.2\textheight]{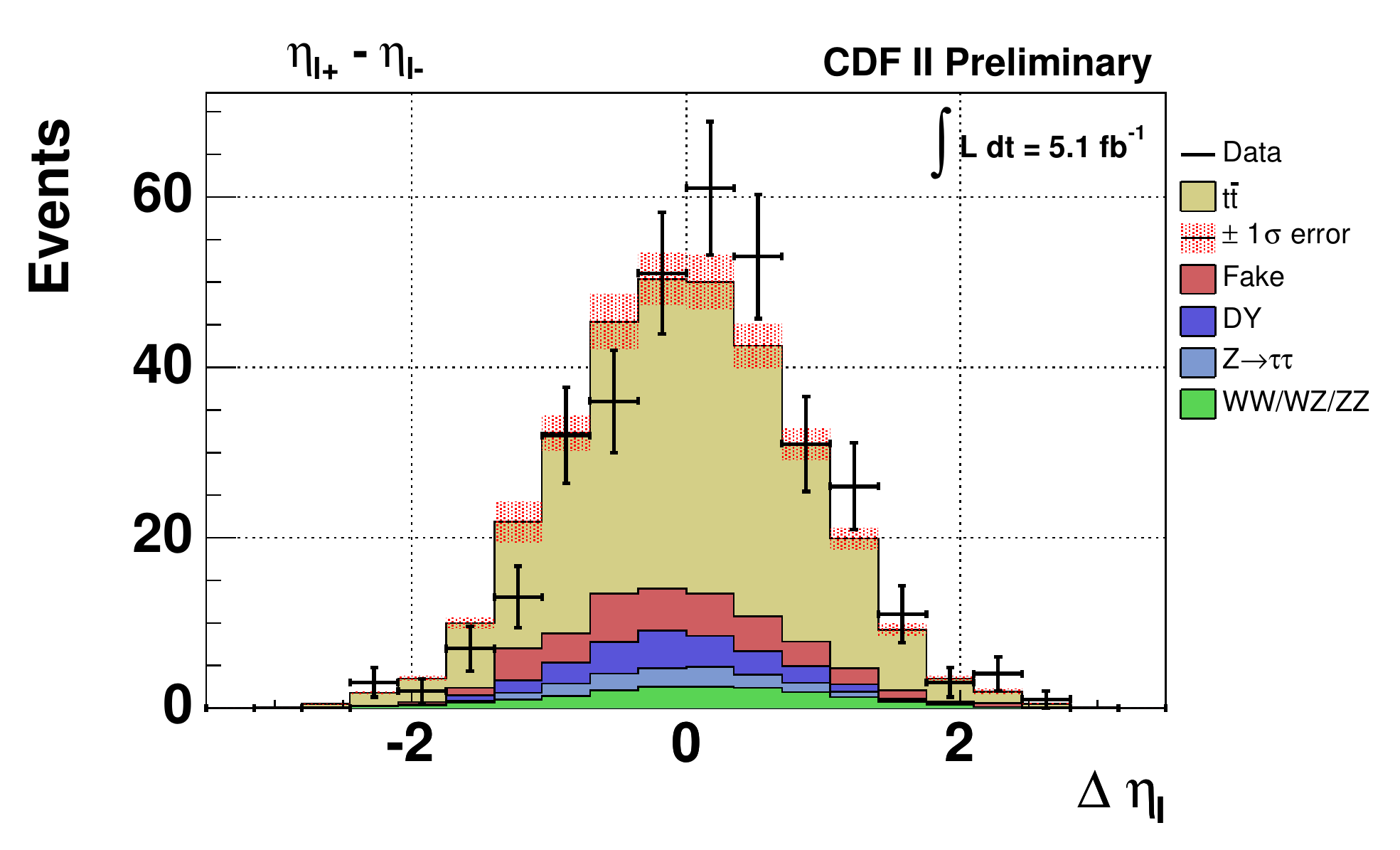} &
        \includegraphics[height=.2\textheight]{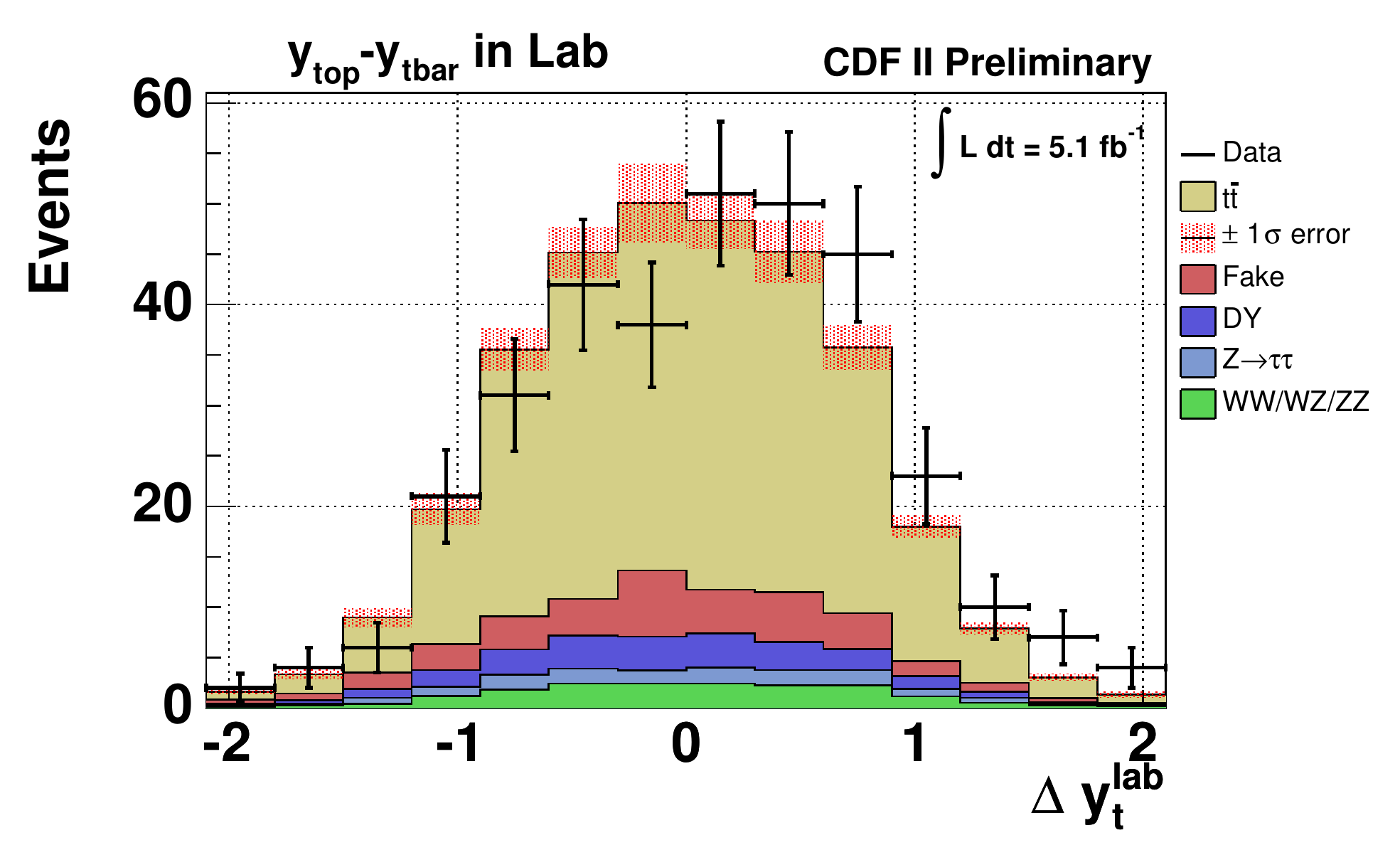} \\
     \end{tabular}
  \caption{Left: The difference of the pseudo-rapidity of the two leptons in the same event.
	   Right: The rapidity difference of the top and anti-top.}
  \label{Deta_ll_dyt}
\end{figure}

To study the asymmetry we plot the pseudo-rapidity difference of 
the two leptons, which is highly correlated with the difference 
of the rapidity of the top and anti-top quarks. This is shown in
FIGURE \ref{Deta_ll_dyt}. The raw asymmetry is calculated to be 
$0.14 \pm 0.05 (stat)$.

The event candidates are reconstructed using a likelihood method. With given
assumed pairing of lepton and jet from the same top decay, top quark mass,
$W$ boson mass, the momenta of the neutrinos can be calculated with the
constraint of missing transverse energy. $P_z^{t \bar{t}}$, $p_T^{t \bar{t}}$,
and $M^{t \bar{t}}$ are calculated and compared to the distributions from 
the MC study. Jet energy and missing transverse energy are also allowed 
to fluctuate within the resolutions. Solution is choosen based on the 
best likelihood value. For the solution choosen rapidities of top and
anti-top quarks are calculated. The difference is shown in the 
FIGURE ~\ref{Deta_ll_dyt}. The raw asymmetry is $0.14 \pm 0.05 (stat)$. 
With background subtraction this becomes $0.21 \pm 0.07 (stat) \pm 0.02 
(background shape)$. Converting this to the parton level, taking into
account the detector acceptance and the reconstruction efficiency, the
true asymmetry is measured to be $0.42 \pm 0.15 (stat) \pm 0.05(sys)$.

\section{Conclusion}

Using 5 fb$^{-1}$ data CDF has done the measurements of the top anti-top pair
production asymmetry study in both lepton plus jets and di-lepton channels. 
Both measurements show significant deviation from the QCD NLO predictions. Many
theories have been proposed to explain it. With more data analyzed in the near
future this study could further strenthen the results and try to identify the
physics that causes this.


\begin{theacknowledgments}
   We thank the Fermilab staff and the technical staffs of the participating
institutions for their vital contributions. This work was supported by the
U.S. Department of Energy and National Science Foundation; the Italian Istituto
Nazionale di Fisica Nucleare; the Ministry of Education, Culture, Sports,
Science and Technology of Japan; the Natural Sciences and Engineering Research
Council of Canada; the National Science Council of the Republic of China; the
Swiss National Science Foundation; the A.P. Sloan Foundation; the Bundesministerium
fur Bildung and Forschung, Germany; the Korean World Class University Program,
the National Research Foundation of Korea; the Science and Technology Facilities
Council and the Royal Society, UK; the Institut National de PHysique Nucleaire et 
Physique des Particules/CNRS; the Russian Foundation for Basic Research; the 
Ministerio de Ciencia e Innovacion, and Programa Consolider-Ingenio 2010, Spain;
the Slovak R\&D Agency; the Academy of Finland; and the Australian Research Council (ARC).
I thank the organizers of the DIS 2011 conference and the Jefferson Lab for providing
such nice chance to discuss various interesting physics topics.
\end{theacknowledgments}



\bibliographystyle{aipprocl} 




\end{document}